\documentclass[prx,twocolumn,amsmath,amssymb]{revtex4}

\usepackage{graphicx}
\usepackage{dcolumn}
\usepackage{bm}
\usepackage{mathrsfs}
\usepackage{subfigure}
\usepackage{natbib}
\usepackage{amsmath}
\usepackage{amssymb}
\usepackage{Jonasmacros}
\usepackage[usenames,dvipsnames,svgnames]{xcolor}
\usepackage{mathptmx}
\usepackage{txfonts}

\begin{document}
	\title{Spin-dependent heat signatures of single-molecule spin dynamics}

	\author{H. Hammar}
	\affiliation{Department of Physics and Astronomy, Uppsala University, Box 530, SE-751 21 Uppsala}

	\author{J. D. Vasquez Jaramillo}
	\affiliation{Department of Physics and Astronomy, Uppsala University, Box 530, SE-751 21 Uppsala}

	\author{J. Fransson}
	\affiliation{Department of Physics and Astronomy, Uppsala University, Box 530, SE-751 21 Uppsala}

	\date{\today}

\begin{abstract}
We investigate transient spin-dependent thermoelectric signatures in a single-molecule magnet under the effect of a time-dependent voltage pulse. We model the system using nonequilibrium Green's functions and a generalized spin equation of motion incorporating the dynamic electronic structure of the molecule. We show that the generated heat current in the system is due to both charge and spin contributions, related to the Peltier and the spin-dependent Peltier effect. There is also a clear signature in the heat current due to the spin dynamics of the single-molecule and a possibility to control the spin-dependent heat currents by bias, tunneling coupling and exchange interaction. A reversal of the net heat transfer in the molecule is found for increasing bias voltage due to the local Zeeman split and we can correlate the net heat transfer with the local anisotropies and dynamic exchange fields in the system.
\end{abstract}

\maketitle

\section{Introduction}

Thermoelectricity and thermodynamics in nanosystems, such as single-molecules and nanojunctions, have been under investigation during recent years \cite{Dubi2011}. Together with the experimental realizations and control of single-molecule magnets (SMMs), and the extension of conventional thermoelectrics to include spin degrees of freedom, has led to the conjunction of spin-dependent thermoelectric effects in nanoscale systems.

Spin-dependent thermoelectricity has been studied in molecular systems and quantum dots \cite{Dubi2009a, Swirkowicz2009, Wang2010, Trocha2012, Weymann2013, Ren2014, Karwacki2013, Misiorny2014, Misiorny2015, Weymann2016, Karwacki2016, Hwang2016, Ramos-Andrade2017, Trocha2017}. Furthermore, other studies involve the effect of time-dependent control on the energy and heat transfer of molecular systems \cite{Crepieux2011,Tagani2012,Liu2012a,Zhou2015,Dare2016} to, e.g., improve the thermoelectric efficiency or design thermal machines. Similarly, electrical and thermal control of, e.g., local interactions and anisotropies in SMMs have been demonstrated both experimentally and theoretically \cite{Loth:2010aa,Khajetoorians2011, Khajetoorians2013, Fahrendorf:2013aa, Heinrich2015, Campbell:2016aa, Natterer2017, Fransson2014a, Hammar2016, Saygun2016a, Jaramillo2017, Jaramillo2018, JaramilloThesis}, thus, defining new realms for engineered driven nanoscale thermoelectric devices containing SMMs.

The discovery of the spin Seebeck and spin Peltier effect led to further increase the interest in spin-dependent thermoelectric effects \cite{Xiao2010, Uchida2010a, Goennenwein2012,Bauer2012, Adachi2013}. It also includes investigations of the spin-dependent Seebeck and Peltier effect where the heat current is coupled to the spin-dependent electron channels in the material. In the context of SMMs, local anisotropies have been suggested to have an effect on the spin-dependent thermoelectric transport properties \cite{Wang2010, Misiorny2014, Misiorny2015}. Experiments of SMMs show thermodynamic signatures of the change of spin configurations in the magnet \cite{Sharples2014}, similar to the effect of spin entropy in bulk materials \cite{Wang2003}.

\begin{figure}[b]
	\centering
	\includegraphics[width=0.75\columnwidth]{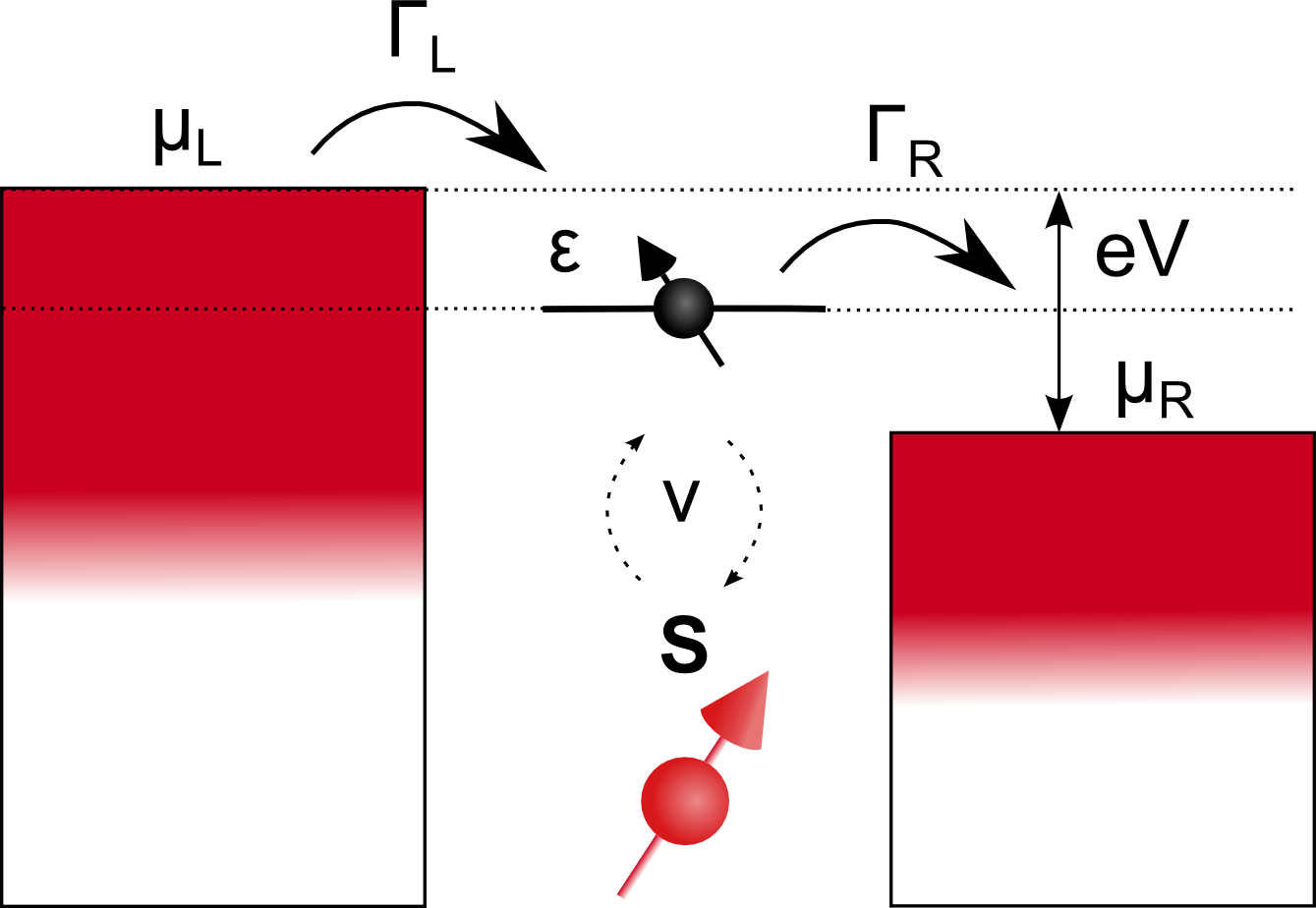}
	\caption{The system studied in this work, consisting of a local magnetic moment coupled to a quantum dot in a tunnel junction between magnetic leads.}
	\label{system}
\end{figure}

Experimental progress in nanoscale systems has made it possible to observe heat using nanothermometry \cite{Pekola2015} and ultrafast spin dynamics with nanoscale resolution \cite{Yoshida2014}. This opens up the possibility to probe and study the relation between heat and spin dynamics in nanoscale systems, specially in SMMs. Theoretically, there has been limited number of attempts to approach the connection between spin dynamics in a SMM and its effect on the heat currents. In this paper we investigate the effect of the spin dynamics in the heat current of a SMM for a system under the influence of a voltage pulse. We model the system as a local magnetic moment coupled to a quantum dot (QD) between two magnetic leads where we apply a pulse across the junction, see Fig. \ref{system}. We show that the generated heat current in the junction can be related to both the transport of charges, i.e., conventional thermoelectric effects, and to the spin transport in the system, i.e., spin-dependent thermoelectric effects. This can, in turn, be related to the spin-dependent Peltier effect and we show that there are signatures of the spin dynamics of the localized magnetic moment in the heat flow.

Our results show a reversal of the net heat transfer for an increased bias voltage due to the local Zeeman split in the molecule, which is influenced by the external magnetic field and the localized spin. The spin dynamics and the corresponding spin-dependent heat transfer can be controlled by both the tunneling coupling, exchange interaction and intrinsic uniaxial anisotropy of the SMM. The change of the parameters result in a significant modulation of the dynamic exchange fields of the SMM and our results show a clear effect of the local anisotropies on the thermoelectric properties of the SMM in accordance with previous studies \cite{Wang2010, Misiorny2014, Misiorny2015}.

Our test bench model represents a single-molecule magnet, for instance $M$-porphyrins and $M$-phthalocyanines where $M$ denotes, e.g., a transition metal element, and which have been experimentally realized and measured \cite{Urdampilleta2011, Vincent2012, Ganzhorn2013, Heinrich2015, Krainov2017, Godfrin2017}. In such compounds we can separate the magnetic molecule into a QD level and a localized magnetic moment \cite{Godfrin2017}. This is justified since the transition metal $d$-levels, which are deeply localized, constitute the localized magnetic moment. The $s$- and $p$-orbitals in the ligands, however, generate the spectral intensity at the highest occupied molecular orbital (HOMO) and lowest unoccupied molecular orbital (LUMO) levels, which is considered as the QD level(s) in our model. Therefore, our model is restricted to large spin moments, for which a classical description is viable, while quantum spins are beyond our approach. We, moreover, assume the QD level to be resonant with the equilibrium chemical potential, hence, avoiding possible Kondo effect that otherwise may occur. While neglecting the local Coulomb repulsion is a severe simplification of the QD description, it is justified since it is typically negligible for the $sp$-orbitals that constitute the conducting levels in the molecular ligands structure. We note though, that including strongly correlated phenomena could have interesting effects on the thermoelectric properties as shown in Ref. \cite{Trocha2012, Weymann2013, Karwacki2013, Weymann2016, Karwacki2016}.

We limit ourselves to discuss heat related to voltage changes between the leads. We do, therefore, not consider effects caused by strong coupling between the system and the bath. Although this limitation might be considered as severe. However, since it yet remains an open question to properly define quantum thermodynamics for strongly coupled systems \cite{Esposito2015, Seifert2016}, we consider this limitation worth testing in this context. Hence, we discard reactance contributions to the energy current caused by the system bath coupling, considered in Ref. \cite{Ludovico2014,Ludovico2016b,Haughian2018}. It is, moreover, still unclear in what sense the energy reactance exhibits itself in response to a sudden on/off-set of the pulse, as well as other shortcomings with the approach that remains to be solved \cite{Esposito2015b, Bruch2016a, Bruch2018}. Therefore, we choose to regard the heat and energy transport as from deep within the leads as motivated in Ref. \cite{Liu2012a}. We note that our treatment might not fully comply with the thermodynamic laws in a consideration of the molecule alone, and that further considerations may also need to include local entropy production in order to account for the transient dynamics \cite{Esposito2015} and the thermodynamics of the localized spin.

The paper is organized as follows. In Sec. II we lay out the theoretical background to the study. This includes the specific model under study, a general background to thermoelectricity in a SMM, heat current with charge- and spin-dependent parts and a brief review of the single-molecule spin dynamics employed in the study. In Sec. III we show the results of our numerical simulations and in Sec. IV we conclude the paper.

\section{Theory}

\subsection{Model system}
We consider a magnetic molecule, embedded in a tunnel junction between metallic leads, comprising a localized magnetic moment ${\bf S}$ coupled via exchange to the highest occupied or lowest unoccupied molecular orbital henceforth referred to as the QD level. The system is shown in Fig. \ref{system} and we define our system Hamiltonian as
\begin{equation}
{\cal H}
=
{\cal H_{{\rm L}}}+{\cal H_{{\rm R}}}+{\cal H_{{\rm T}}}+{\cal H_{{\rm QD}}}+{\cal H_{{\rm S}}}
.
\end{equation}
Here, ${\cal H}_{\chi}=\sum_{\bfk\sigma\in\chi}(\varepsilon_{\bfk\chi}-\mu_{\chi}(t))c_{\bfk\chi\sigma}^{\dagger}c_{\bfk\chi\sigma},$
is the Hamiltonian for the left ($\chi=L$) or right ($\chi=R$) lead, where $c_{\bfk\chi\sigma}^{\dagger}$ ($c_{\bfk\chi\sigma}$) creates (annihilates) an electron in the lead $\chi$ with energy $\varepsilon_{\bfk\chi}$, momentum \textbf{k}, and spin $\sigma=\up,\down$, while $\mu_{\chi}(t)$ denotes the chemical potential such that the voltage $V(t)$ across the junction is defined by $eV(t) = \mu_{L}(t)-\mu_{R}(t)$.
Tunneling between the leads and the QD level is described by ${\cal H}_{T}={\cal H}_{TL}+{\cal H}_{TR}$, where $
{\cal H}_{T\chi}=T_{\chi}\sum_{\bfk\sigma\in\chi}c_{\mathbf{k\chi\sigma}}^{\dagger}d_{\sigma}+H.c.$.
The single-level QD is represented by ${\cal H}_{QD}=\sum_{\sigma}\varepsilon_{\sigma}d_{\sigma}^{\dagger}d_{\sigma}$, where $d_{\sigma}^{\dagger}$ ($d_{\sigma}$) creates (annihilates) an electron in the QD with energy $\varepsilon_{\sigma} = \varepsilon_{0} + g\mu_{B} B^\text{ext} \sigma^{z}_{\sigma\sigma}/2$ and spin $\sigma$, depending on the external magnetic field $\bfB^\text{ext}=B^\text{ext}\hat{\bf z}$, where g is the gyromagnetic ratio and $\mu_{B}$ the Bohr magneton. The energy of the local spin is described by ${\cal H}_{\rm S}=-g\mu_B\bfS\cdot\bfB^\text{ext}-v\mathbf{s}\cdot\mathbf{S}-DS_{z}^2$, where $v$ is the exchange integral between the localized and delocalized electrons, the electron spin is denoted \textbf{$\mathbf{s}=\psi^{\dagger}\boldsymbol{\sigma}\psi/2$} in terms of the spinor $\psi=(d_\up,\ d_\down)$, $\boldsymbol{\sigma}$ is the vector of Pauli matrices and D is an intrinsic uniaxial anisotropy field in the magnetic molecule.

The dynamical QD electronic structure is calculated by using nonequilibrium Green's functions (GF), defined on the Keldysh contour ${\bf G}(t,t')=\theta(t-t'){\bf G}^>(t,t')+\theta(t'-t){\bf G}^<(t,t')$. We take into account the back action from the local spin dynamics by perturbation theory, expanding to first order in the time-dependent expectation value of the spin according to

\begin{align}
\mathbf{G}(t,t')=&
{\bf g}(t,t')-v\oint_C {\bf g}(t,\tau)\left\langle \mathbf{S}(\tau)\right\rangle \mathbf{\cdot\boldsymbol{\sigma}}{\bf g}(\tau,t')d\tau.
\end{align}
Here, ${\bf g}(t,t')$ is the bare QD GF defined as a $2\times2$-matrix in spin space. It is defined by the equation
\begin{align}
(i\partial_{t}-\boldsymbol{\epsilon}){\bf g}(t,t')=&\delta(t-t')\sigma^0+\int\boldsymbol{\Sigma}(t,\tau){\bf g}(\tau,t')d\tau
,
\end{align}
where $\boldsymbol{\epsilon}$ is the matrix of spin-dependent energy levels of the QD and the self-energy is defined as  $\boldsymbol{\Sigma}(t,t')=\sum_\chi\sum_{\bfk\in\chi}|T_\chi|^2\textbf{g}_{\bfk\sigma}(t,t')$, where $\textbf{g}_{\bfk\sigma}(t,t')$ is the GF for electrons in the lead.

Solving for the magnetic lead GF, the self-energies can be expressed as
\begin{align}
\boldsymbol{\Sigma}^{</>}(t,t')=&
(\pm i)\sum_\chi\boldsymbol{\Gamma}^\chi \int f_{\chi}(\pm\omega)e^{-i\omega(t-t')+i\int_{t'}^{t}\mu_{\chi}(\tau)d\tau}\frac{d\omega}{2\pi},
\label{selfenergy}
\end{align}
where we have introduced the coupling matrix $\bfGamma^\chi=\Gamma^\chi_0\sigma^0+\bfGamma^\chi_1\cdot\bfsigma$ and defined the tunneling couplings $\Gamma_\sigma^\chi=2|T_\chi|^2\sum_{\bfk\in\chi}\delta(\omega-\leade{\bfk})$, $\Gamma^{\chi}_{0}=\sum_{\sigma}\Gamma^{\chi}_{\sigma}$ and $\boldsymbol{\Gamma}^{\chi}_1=\sum_{\sigma}\sigma_{\sigma\sigma}^{z}\Gamma^{\chi}_{\sigma}\mathbf{\hat{z}}$ using the wide-band limit. By introducing the spin-polarization in the leads $p_{\chi}\in\left[ -1,1\right]$, such that $\Gamma^{\chi}_{\sigma}=\Gamma_{0}^{\chi}(1+\sigma_{\sigma\sigma}^{z}p_{\chi})/2$, we can write $\boldsymbol{\Gamma}^{\chi}_1=p_{\chi}\Gamma^{\chi}_{0}\mathbf{\hat{z}}$.
Using similar notation we can write the lesser/greater GF and self-energies as $\bfG^{</>}=\bfG_0^{</>}\sigma^0+\bfsigma\cdot\bfG_1^{</>}$ and $\bfSigma^{</>} =\Sigma_0^{</>}\sigma^0+\bfSigma_1^{</>}\cdot\bfsigma$. In Appendix \ref{app-NEGF} we show the final form of the lesser/greater GF expressed in its charge and spin components. We refer to Ref. \cite{Hammar2016} for more details. The self-energy carries the information of an on-set of the voltage in the system to the time integration of the chemical potential for each lead, i.e., $i\int_{t'}^{t}\mu_{\chi}(\tau)d\tau$. Thus, this initiates the dynamics in the system and carries the information of the pulse.

\subsection{Thermoelectricity in a single-molecule magnet}
Thermoelectric effects connect heat, charge and spin biases with the currents. Conventional thermoelectric effects relate heat and charge currents through the Seebeck and Peltier effects, and recent progress has extended these concepts to spin-dependent counterparts. In this article we focus on the Peltier and the spin-dependent Peltier effects.

The normal Peltier effect is the heat current response to a charge current, where heat is carried by the charges. In literature, there is a distinction between two types of spin related Peltier effects. One type, the spin-dependent Peltier effect, where the heat transfer is generated by the spin current that results from a spin-imbalance in the charge current. The second type, the spin Peltier effect, is a collective phenomena that emerges even in absence of charge transfer, i.e, $I_C = 0$.

In a SMM that contains a local magnetic moment, there can be both spin-independent and spin-dependent thermoelectric effects. The normal Peltier effect arises in presence of charge currents in the system. However, local electron scattering off the magnetic moment can give rise to the spin-dependent Peltier effect, since a net spin current may be generated by the scattering. Furthermore, a finite spin bias between the leads can generate a pure spin current, that is, the spin Peltier effect \cite{Wang2010, Ren2014, Misiorny2014, Misiorny2015}. Here, while we do not include spin biases in our discussions, our calculated spin-dependent heat transfers only pertain to the spin-dependent Peltier effect.

\subsection{Heat and energy currents}
The properties of the QD are probed by means of the heat and energy currents flowing through the system. In this way, the goal is to pick up signatures of the spin dynamics in the thermoelectric transport properties. We start by defining the particle, energy and heat current, $I^N$, $I^E$ and $I^Q$, respectively. Accordingly, we define

\begin{subequations}
	\begin{align}
	I^N_\chi(t)=&
	\dt\sum_{\bfk\sigma\in \chi}\av{n_{\bfk\sigma}}
	= \frac{i}{\hbar}\sum_{\textbf{k}\sigma}\left\langle \left[c^{\dagger}_{\textbf{k}\sigma\chi}c_{\textbf{k}\sigma\chi},\mathcal{H}\right] \right\rangle
	,
	\\
	I^E_\chi(t)=&
	\dt\av{\mathcal{H}_{\chi}}= \frac{i}{\hbar}\sum_{\textbf{k}\sigma} \varepsilon_{\textbf{k}\sigma}\left\langle \left[c^{\dagger}_{\textbf{k}\sigma\chi}c_{\textbf{k}\sigma\chi},\mathcal{H}\right] \right\rangle
	,
	\\
	I_{\chi}^{Q} = & I_{\chi}^{E} - \mu_{\chi}I_{\chi}^{N}.
	\label{heatcurrent}
	\end{align}
\end{subequations}

Using standard methods we can write the particle current as

\begin{align}
I_{\chi}^{N}(t) = & \frac{2}{\hbar}\operatorname{sp}\int_{-\infty}^{t}\left(\boldsymbol{\Sigma}_{\chi}^{>}(t,t')\mathbf{G}^{<}(t',t) \right.\nonumber \\
& \left. +\boldsymbol{\Sigma}_{\chi}^{<}(t,t')\mathbf{G}^{>}(t',t)\right)dt'\label{particlecurrent}
,
\end{align}
where ${\rm sp}$ denotes the trace over spin-1/2 space.
Using the generic separation of a matrix $\bfA=A_0\sigma^0+\bfsigma\cdot\bfA_1$, we partition the current into a spin-independent and spin-dependent part according to $I_{\chi}^{N}(t)  =  I_{0\chi}^{N}(t)+I_{1\chi}^{N}(t)$, where

\begin{subequations}
	\begin{align}
	I_{0\chi}^{N}(t)=&
	\frac{4}{\hbar}
	\int_{-\infty}^t
	\Bigl(
	\Sigma_{0\chi}^{>}G_0^<
	+
	\Sigma_{0\chi}^{<}G_0^>
	\Bigr)
	dt'
	,
	\label{particlecurrent0}
	\\
	I_{1\chi}^{N}(t)=&
	\frac{4}{\hbar}
	\int_{-\infty}^t
	\Bigl(
	\boldsymbol{\Sigma}_{1\chi}^{>}\cdot\bfG_1^<
	+
	\boldsymbol{\Sigma}_{1\chi}^<\cdot\bfG_1^>
	\Bigr)
	dt'
	.
	\label{particlecurrent1}
	\end{align}
\end{subequations}
Analogously, the spin-independent and spin-dependent parts of the energy current become
\begin{subequations}
	\begin{align}
	I_{0\chi}^{E}(t)=&
	\frac{4}{\hbar}
	\int_{-\infty}^t
	\Bigl(
	\Sigma_{E0\chi}^{>}G_0^<
	+
	\Sigma_{E0\chi}^{<}G_0^>
	\Bigr)
	dt'
	,
	\label{energycurrent0}
	\\
	I_{1\chi}^{E}(t)=&
	\frac{4}{\hbar}
	\int_{-\infty}^t
	\Bigl(
	\boldsymbol{\Sigma}_{E1\chi}^{>}\cdot\bfG_1^<
	+
	\boldsymbol{\Sigma}_{E1\chi}^<\cdot\bfG_1^>
	\Bigr)
	dt'
	.
	\label{energycurrent1}
	\end{align}
\end{subequations}
Here, we have defined the energy self-energy as $
\Sigma^{</>}_{E\chi\sigma}(t,t') =\sum_{\mathbf{k} \in \chi} \varepsilon_{\textbf{k}\chi\sigma} |T_\chi|^2 g^{</>}_{\bfk\sigma}(t,t')$ and write $\bfSigma_E^{</>} =\Sigma_{E0}^{</>}\sigma^0+\bfSigma_{E1}^{</>}\cdot\bfsigma$.

Using the decomposition scheme in terms of charge and spin components, that is, $I_{\chi}^{Q}(t)  =  I_{0\chi}^{Q}(t)+I_{1\chi}^{Q}(t)$, we can distinguish between the two Peltier effects. The contributions related to the Peltier effect are contained in $I_{0\chi}^{Q}(t)$, whereas $I_{1\chi}^{Q}(t)$ contains the contributions to the corresponding the spin-dependent Peltier effect.

\subsection{Single-molecule spin dynamics}
The local spin dynamics is calculated using our previously developed generalized spin-equation of motion (SEOM) \cite{Hammar2016}

\begin{align}
\dot{\mathbf{S}}(t) = &\mathbf{S}(t)\times\left( -g\mu_{B}\mathbf{B}^{\mathrm{eff}}_{0}(t)
+\frac{1}{e}\int (\mathbb{J}(t,t')+\mathbb{D})\cdot\mathbf{S}(t')dt'\right).
\label{spinequationofmotion}
\end{align}

Here, $\mathbf{B}^\text{eff}_{0}(t)$ is the effective magnetic field acting on the spin, defined by $g\mu_{B}\mathbf{B}^{\mathrm{eff}}_{0}(t)=g\mu_{B}\textbf{B}^{\mathrm{ext}}+v\bfm(t)-\int\textbf{j}(t,t')dt'/e$, where the second contribution is the local magnetic occupation, defined as $\bfm (t) = \left\langle \mathbf{s}(t) \right\rangle = \left\langle \psi(t)^\dagger \bfsigma \psi(t)\right\rangle/2=\text{Im sp }{\bfsigma}\textbf{G}^<(t,t)/2$, and the third term is the internal magnetic field due to the electron flow. The field $\mathbb{J}(t,t')$ is the dynamical exchange coupling between spins at different times and $\mathbb{D} = D \mathbf{\hat{z}\hat{z}}$ is due to the intrinsic uniaxial anisotropy.

The generalized SEOM makes use of the Born-Oppenheimer approximation which is motivated as the energy scales of single molecule magnets are in meV which results in spin dynamics of picoseconds. This is orders of magnitudes greater than the recombination time-scales of the electrons in the junction in the orders of femtoseconds. We also remark that despite the semi-classical nature of the generalized SEOM, it incorporates the underlying quantum nature of the junction through the dynamical fields ${\bf j}$ and $\mathbb{J}$. This is especially important in the transient regime, where the classical Landau-Lifshitz-Gilbert equation is incapable to provide an adequate description of the dynamics \cite{Hammar2017}. The treatment goes beyond the adiabatic limit considered in previous works on SMM spin dynamics, while still containing important attributes as dissipative fields and spin-transfer torques.

The internal magnetic field due to the electron flow is defined as
$\textbf{j}(t,t')=iev\theta(t-t')\av{\com{s^{(0)}(t)}{\mathbf{s}(t')}}$.The on-site energy distribution is represented by $\text{s}^{(0)}=\sum_{\sigma}\varepsilon_{\sigma}d_{\sigma}^{\dagger}d_{\sigma}$. The two-electron propagator is, here, approximated by decoupling into single electron nonequilibrium GF according to
\begin{eqnarray}
\mathbf{j}(t,t')&\approx&
iev\theta(t-t'){\rm sp}\boldsymbol{\epsilon}
\Bigl(
\mathbf{G}^{<}(t',t)\mathbf{\boldsymbol{\sigma}G}^{>}(t,t')
\nonumber\\&&
-\mathbf{G}^{>}(t',t)\mathbf{\boldsymbol{\sigma}G}^{<}(t,t')
\Bigr),
\label{currentMF}
\end{eqnarray}
where $\boldsymbol{\epsilon}={\rm diag}\lbrace \varepsilon_\uparrow\ \varepsilon_\downarrow\rbrace$. This internal field mediates both the magnetic field generated by the charge flow as well as the effect of the external magnetic field causing the Zeeman split in the QD.

The spin susceptibility tensor $\mathbb{J}(t,t')=i2ev^2\theta(t-t')\av{\com{\bfs(t)}{\bfs(t')}}$ mediates the interactions between the localized magnetic moment at the times $t$ and $t'$. Decoupling $\mathbb{J}$ into single electron GFs, we can write
\begin{align}
\mathbb{J}(t,t')\approx&
\frac{ie}{2}v^{2}\theta(t-t'){\rm sp}\boldsymbol{\sigma}
\Bigl(
\mathbf{G}^{<}(t',t)\mathbf{\boldsymbol{\sigma}G}^{>}(t,t')
\nonumber\\&
-\mathbf{G}^{>}(t',t)\mathbf{\boldsymbol{\sigma}G}^{<}(t,t')
\Bigr).
\label{spinsusceptibilty}
\end{align}
This current-mediated interaction can be decomposed into an isotropic Heisenberg interaction $J_H$ and the anisotropic Dzyaloshinski-Moriya (DM) ${\bf D}$ and Ising $\mathbb{I}$ interactions \cite{Fransson2014a,Hammar2016}. Applying this decomposition to the second term in Eq. \eqref{spinequationofmotion} gives
\begin{align}
\mathbf{S}(t)\times\mathbb{J}(t,t')\cdot\mathbf{S}(t')=& 
J_{H}(t,t')\mathbf{S}(t)\times\mathbf{S}(t')
\nonumber\\&
+\mathbf{S}(t)\times\mathbb{I}(t,t')\cdot\mathbf{S}(t')
\nonumber\\&
-\mathbf{S}(t)\times\mathbf{D}(t,t')\times\mathbf{S}(t').
\label{decomposition}
\end{align}
The interactions $J_H$, ${\bf D}$, and $\mathbb{I}$ can be written in terms of the single electron GF \textbf{G}, which is detailed in Appendix \ref{app-Interactions}.

It has previously been conjectured that the DM interaction corresponds to the current induced spin-transfer torque \cite{Hammar2018}, which here is essential for the switching of the spin. Furthermore, the decomposition of $\mathbb{J}$ enables us to individually calculate the local anisotropies and energy landscape of the SMM. The Ising interaction provides, for instance, a dynamical anisotropy for the molecular spin. Specifically, $\mathbb{I}_{zz}$ is interpreted as a dynamical uniaxial anisotropy. We emphasize that these dynamical anisotropies emerge from the interactions between the electron currents and the local spin moment, and constitutes a dynamical addition to the intrinsic static anisotropy $D$.

We remark that the generalized SEOM does not include the dipolar and quadrupolar fields created by the ferromagnetic leads discussed in Ref. \cite{Misiorny2013}. We note, however, that these fields merely renormalize the external magnetic field and intrinsic anisotropy and can, thereby, be omitted without loss of generality.

\begin{figure}[b]
	\centering
	\includegraphics[width=\columnwidth]{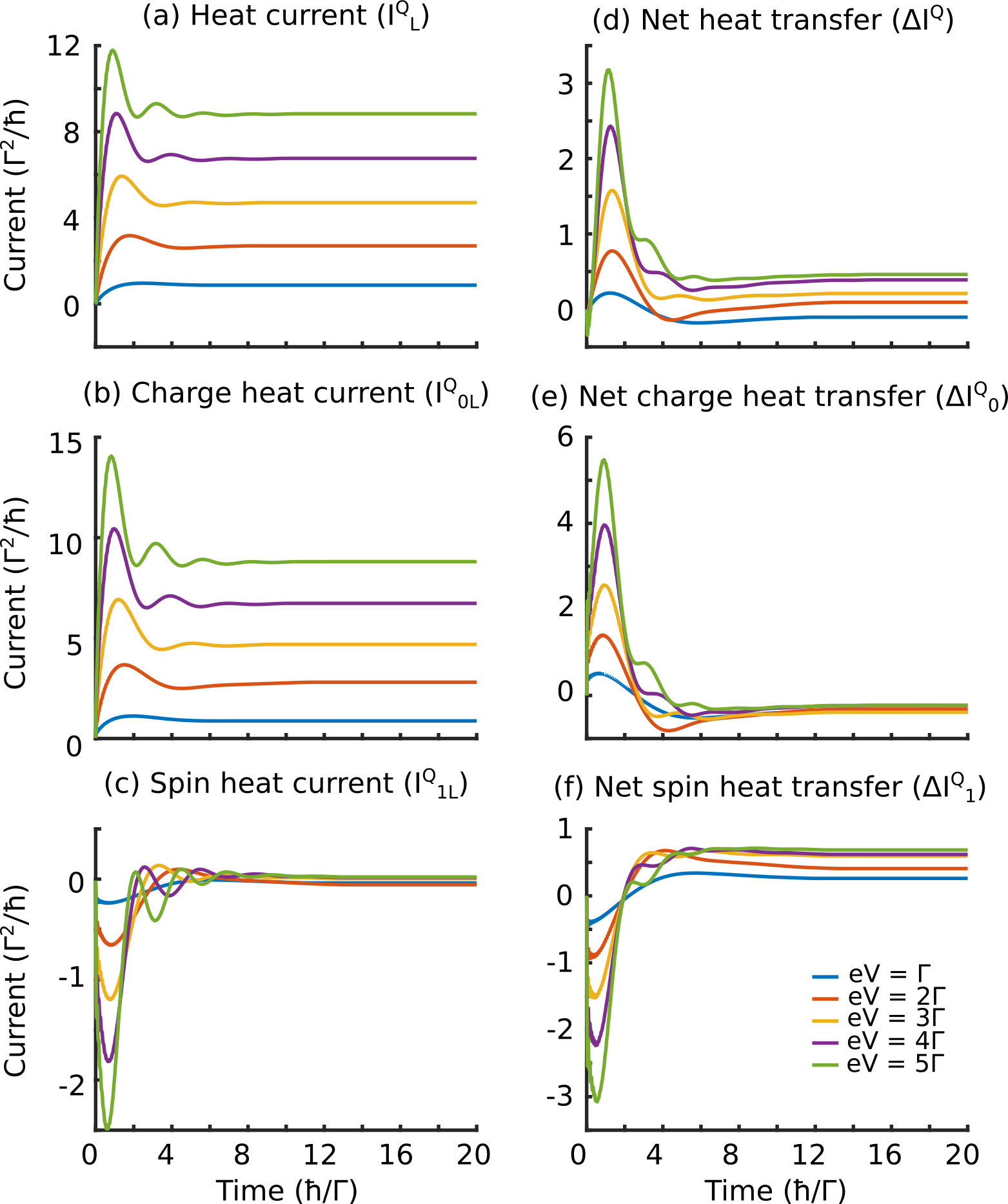}
	\caption{The evolution of the heat currents and net heat transfer in the SMM due to the on-set of a bias voltage for different bias voltages V. First column indicates (a) the heat current flowing from the left lead $I^Q_L$, split into its (b) charge $I^Q_{0L}$ and (c) spin $I^Q_{1L}$ components. Second column indicates (a) the net heat transfer $\Delta I^Q$ split into its (b) charge $\Delta I^Q_{0}$ and (c) spin $\Delta I^Q_{1}$ components. Here, we used $t_0 = 0$, $\Gamma_0 = \Gamma$, $v = \Gamma/2$, $p_L = p_R = 0.5$, $D = 0$, T = 0.0862 $\Gamma/\text{k}_{B}$ and B$^{ext}$ = 0.1158 $\Gamma/\text{g}\mu_{B}$.}
	\label{biasresults}
\end{figure}

\begin{figure*}[t]
	\centering
	\includegraphics[width=0.75\textwidth]{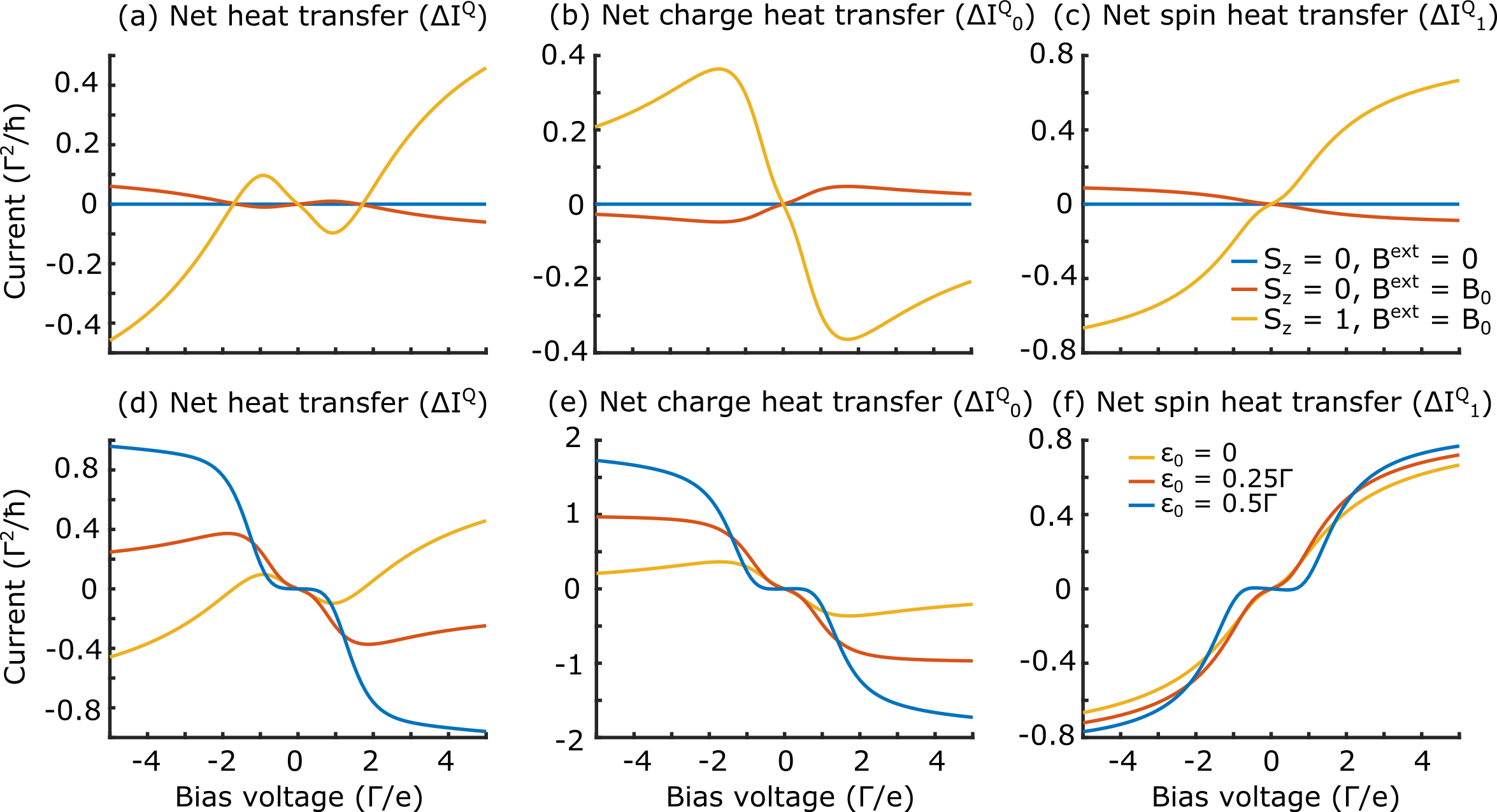}
	\caption{First row represents the stationary solution for the (a) net heat transfer, split into into its (b) charge and (c) spin components for different magnetic fields and values of the spin where $\text{B}_0$ = 0.1158 $\Gamma/\text{g}\mu_{B}$. Second row represents the stationary solution for the (d) net heat transfer, split into into its (e) charge and (f) spin components for different QD energies $\varepsilon_0$ with a finite magnetic field, B$^{ext}$ = $\text{B}_0$, and a finite spin $\text{S}_z$ = 1. Note that the yellow line is the same in both rows. Other parameters as in Fig. \ref{biasresults}.}
	\label{stationaryresults}
\end{figure*}

\section{Results}

\begin{figure}
	\centering
	\includegraphics[width=\columnwidth]{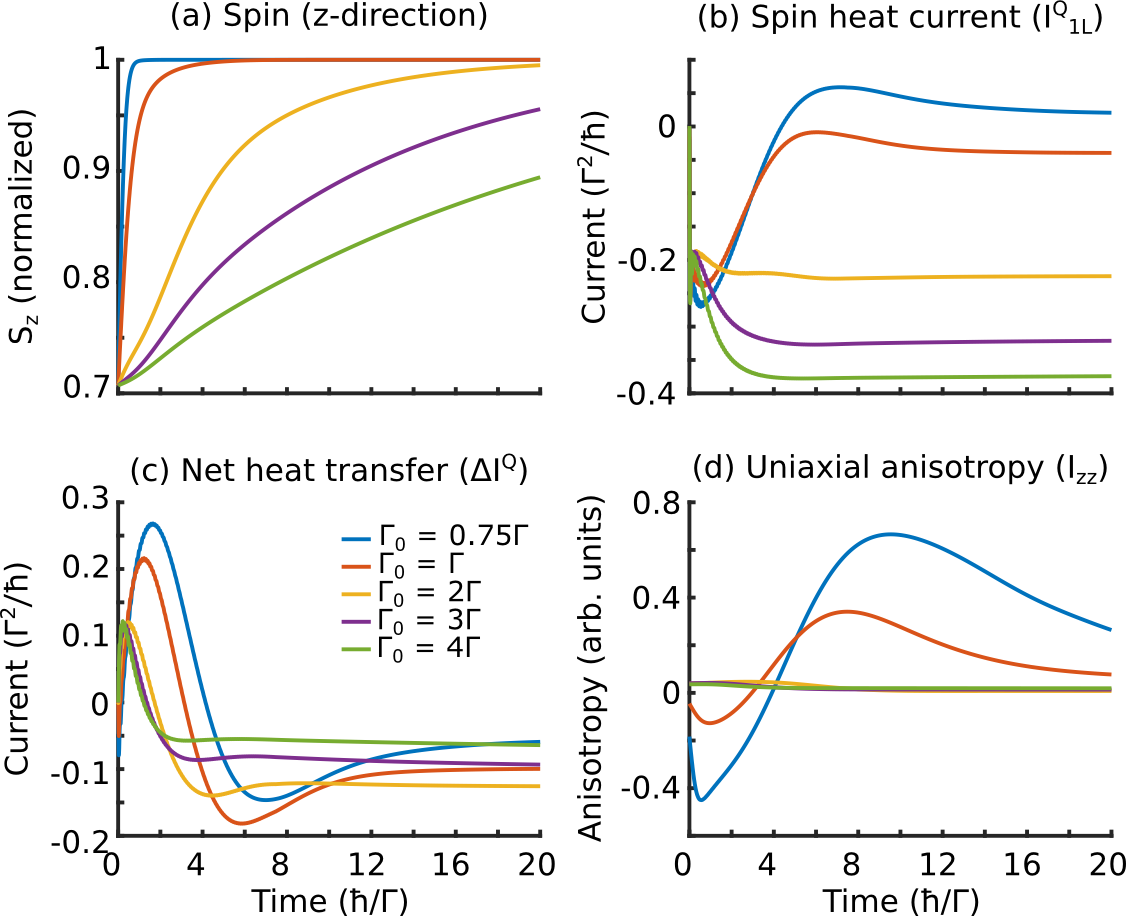}
	\caption{The evolution of (a) the spin, (b) the spin-dependent heat current, (c) the net heat transfer and (d) uniaxial anisotropy part of the Ising interaction for different tunneling coupling $\Gamma_0$. Other parameters as in Fig. \ref{biasresults}.}
	\label{gammaresults}
\end{figure}

\begin{figure*}
	\centering
	\includegraphics[width=0.75\textwidth]{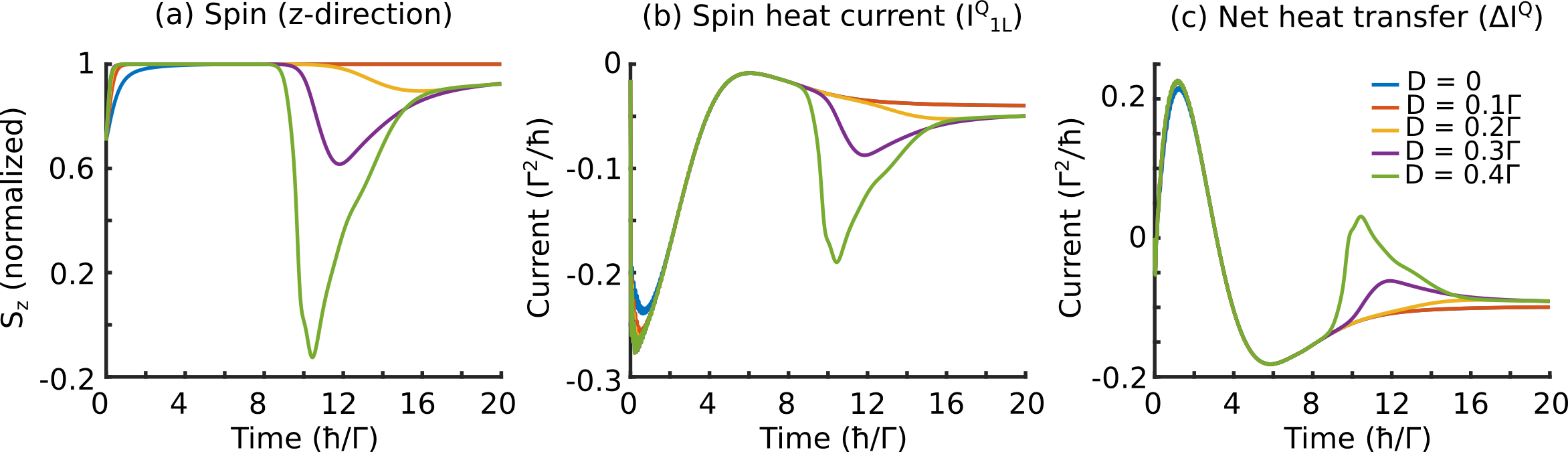}
	\caption{The evolution of (a) the spin, (b) the spin-dependent heat current, (c) the spin-dependent net heat transfer and (d) uniaxial anisotropy part of the Ising interaction for different intrinsic uniaxial anisotropy D. Other parameters as in Fig. \ref{biasresults}.}
	\label{anisotropyresults}
\end{figure*}

\begin{figure}
	\centering
	\includegraphics[width=\columnwidth]{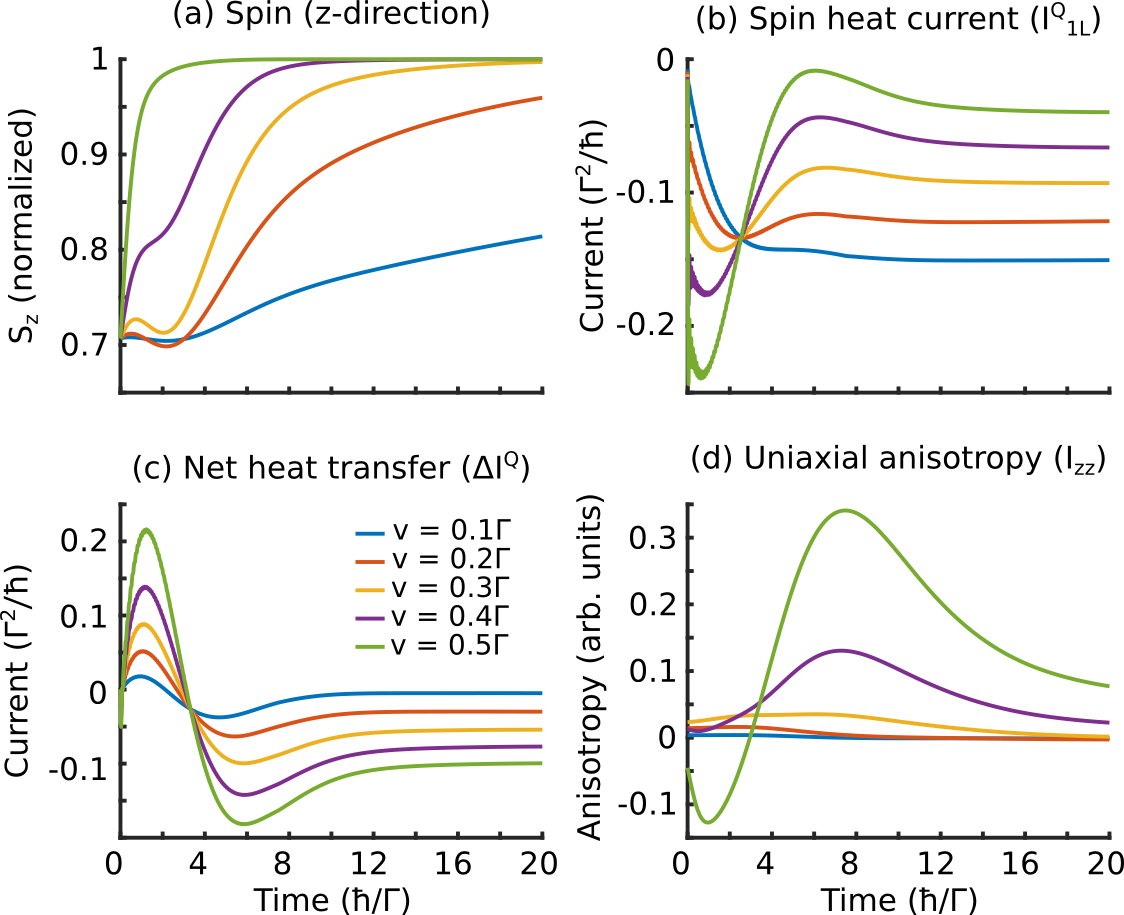}
	\caption{The evolution of (a) the spin, (b) the spin-dependent heat current, (c) the spin-dependent net heat transfer and (d) uniaxial anisotropy part of the Ising interaction for different exchange coupling $v$. Other parameters as in Fig. \ref{biasresults}.}
	\label{exchangeresults}
\end{figure}

We simulate the results for an applied time-dependent bias voltage initiating the dynamical evolution of the localized spin. At a time $t_{0}$ there is a sudden on-set of a bias voltage V that introduces a chemical potential $\mu_{L}(t) = -\mu_{R}(t) = eV\theta(t-t_{0})/2$ to the leads. In order to have a finite spin-dependent heat current we use magnetic leads, i.e., a finite $\boldsymbol{\Gamma}^\chi_{1}$, by setting $p_L = p_R = 0.5$. Before the on-set of the voltage bias, the local spin is subject to the static external magnetic field $\mathbf{B}^{ext}=B^{ext}\hat{\mathbf{z}}$, giving $S_{x}=S_{xy}\sin\omega_{L}t$, $S_{y}=S_{xy}\cos\omega_{L}t$ and $S_{z}=S_{z}$ where $S_{xy}^2=S_x^2+S_y^2$, whereas $\left|S\right|^{2}=S_{xy}^{2}+S_{z}^{2}$ and $\omega_{L}=g\mu_{B}\left|B^{ext}\right|$, and we assume an initial polar angle of $\pi/4$. We also set the temperature of the leads to be the same in order to have pure Peltier contribution in the current and no Fourier heat transfer. We represent the quantities in terms of the model parameter $\Gamma$, which represent the tunneling coupling $\Gamma_0$ in all figures except Fig. \ref{gammaresults} where the tunneling coupling is varied. As motivated in the introduction we are considering $M$-porphyrins and $M$-phthalocyanines where $M$ denotes, e.g., a transition metal element. Here, the typical energies scales are of the orders of meV when it comes to tunneling coupling, exchange interaction and anisotropy \cite{Urdampilleta2011, Vincent2012, Ganzhorn2013, Heinrich2015, Krainov2017, Godfrin2017} . A tunneling coupling of $\Gamma = 1$ meV will result in that the parameters in Fig. \ref{biasresults} becomes $v = 0.5$ meV, T = 1 K, B = 1 T and $\hbar/\Gamma = 0.658$ ps.

The heat current is simulated for different bias voltages, see Fig. \ref{biasresults}. Here, the first column indicates the heat flow from the left lead into the QD. The full heat current is shown in Fig. \ref{biasresults}(a) and is then split into its (b) charge- and (c) spin-dependent components. As we can see, the majority of the contribution due to the charge heat flow is positive. The heat is thus carried by the charge current travelling from the left to the right lead due to the applied bias voltage. The spin-dependent heat current, shown in Fig. \ref{biasresults}(c), is negative, thus counteracts the charge heat current. This contribution quickly vanish for longer times and the main contribution to the heat flow is due to the charge flow in the system. It is although clear that a spin-dependent Peltier effect is of importance in the transient regime and will create a spin-dependent heat flow.

The second column of Fig. \ref{biasresults} shows the net heat transfer in the system. We define it as $\Delta I^{Q} = I_{L}^{Q} - I_{R}^{Q}$, thus, requiring the net contribution due to particle current to be zero in the steady state. It can easily be motivated as $I_{L}^{Q} - I_{R}^{Q} = I_{L}^{E} - \mu_{L}I_{L}^{N} - ( I_{R}^{E} - \mu_{R}I_{R}^{N}) = I_{L}^{E} - I_{R}^{E} - \mu_{L}(I_{L}^{N} + I_{R}^{N})$, where $I_{L}^{N} + I_{R}^{N}$ is zero in the steady-state regime because of particle conservation. Here, we used the fact that $\mu_{R} = -\mu_{L}$ in the present problem. As a result, the long term behavior of the net heat transfer will be dominated by the energy current. The net heat transfer describes the energy transfer through the junction. Note that the net heat transfer considered here is only due to the heat generated by the Peltier effect and spin-dependent Peltier effect. We are not considering Fourier heat and Joule heating as we have no temperature difference and do not include any dissipative mechanism in the molecule. As stated in the introduction we are not considering the full thermodynamics of the dynamic SMM and it could be contributions due to energy stored or released in the molecule.

Fig. \ref{biasresults}(d) shows the net heat transfer due to (e) charge and (f) spin components. As can be seen in Fig. \ref{biasresults}(e)-(f), while the net heat transfer due to the charge flow is initially positive and then negative, the opposite appears for the spin-dependent component. Increasing the bias voltage changes the strength of the different contributions and the magnitude of the spin-dependent net heat transfer increases, creating a net reversal in the heat transfer in the stationary limit.

The net reversal in the heat transfer can be explained using time-independent calculations and is due to the Zeeman splitting of the QD in the junction. As shown in Fig. \ref{stationaryresults}(a), there is no net heat transfer without any external magnetic field or local magnetic moment. Adding a field or a finite local magnetic moment, creates a Zeeman split in the QD, hence, creating an energy difference in the QD states. This energy difference results in a finite energy current, hence, a net heat transfer through the molecule. This can, in turn, be split into its (b) charge and (c) spin components. As seen in Fig. \ref{stationaryresults}(a) - (c), the external magnetic field and the localized spin contribute with opposite signs.
A positive magnetic field, in the z-direction, creates a Zeeman split of the molecular orbitals according to $\varepsilon_\sigma = \varepsilon_0 +\sigma^z_{\sigma\sigma} g\mu_{B} B^\text{ext}/2$. For the localized spin moment, however, the order of the energies is the opposite due to the ferromagnetic exchange interaction. Hence, for a spin moment in the z-direction, there is a lower energy associated with the spin-up electrons than for the spin-down.
It is important to note that for a dynamic spin moment, as in Fig. \ref{biasresults}, the total Zeeman split is dynamic, hence, contributing to the dynamic behavior of the net heat transfer. Furthermore, we can observe that the net heat transfer can be tuned by changing the energy of the QD by applying a gate voltage. As shown in \ref{stationaryresults}(d), the net heat transfer through the junction is inverted when the energy level of the QD is increased. This inversion can be attributed to the increased contribution from the charge component of the heat transfer, Fig. \ref{stationaryresults}(e), which overcomes the spin component, Fig. \ref{stationaryresults}(f).
The reason for the increased charge component in the heat transport, compared to the corresponding spin component, is that the gate voltage pushes the molecular level away from the equilibrium chemical potential, that is $\dote{0}-\mu\neq0$. This leads to a broken left-right symmetry of the junction, which causes the energy difference between the chemical potential of the left lead $\mu_L$ and the QD level $\dote{0}$ to becomes smaller than the corresponding difference between the right lead and the QD level. This imbalance between the leads, hence, generates a net heat transfer.

Next, we look at the dependence on the tunneling coupling $\Gamma_0$. This can be related to the coupling between a STM tip and a molecule, which determines the rate of damping of the system and the anisotropy of the molecule \cite{Heinrich2015}. In Fig. \ref{gammaresults}, the results are shown for different couplings $\Gamma_0$, where Fig. \ref{gammaresults}(a) shows the resulting spin dynamics for the z-projection of the spin. As can be seen in the figure, small couplings result in faster dynamics than large couplings. The effect in the spin-dependent heat current, shown in Fig. \ref{gammaresults}(b), is an increase in the negative spin-dependent contribution for higher exchange couplings. In the case of the net heat transfer, shown in Fig. \ref{gammaresults}(c), increasing the tunneling coupling first gives an increase in the net heat transfer. Then, for higher exchange couplings, the net heat transfer tends towards zero.

The damping behavior in the spin dynamics of the SMM and coupling dependent change of the heat currents is an interplay between the tunneling coupling and the local exchange within the SMM. The increased negative contribution, Fig. \ref{gammaresults}(b), for higher coupling is due to a decreased spin current in the junction as the tunneling starts to dominate over the local exchange. 
Furthermore, as shown in Eq. \eqref{decomposition} the dynamic local exchange field can be decomposed into isotropic Heisenberg interaction and anisotropic DM and Ising interactions. Small tunneling couplings $\Gamma_0$ in comparison to the exchange coupling $v$, creates larger dynamic anisotropies. In Fig. \ref{gammaresults}(d), the time-dependent evolution of the dynamical uniaxial anisotropy ($\mathbb{I}_{zz}$) is shown for the different tunneling couplings. As seen, the local anisotropies vary longly for small tunneling couplings. In the case of larger couplings, the uniaxial anisotropies almost vanish. Thus, there is a large change of the local spin environment due to the tunneling coupling because of the interplay between the tunneling and local exchange, creating a significant change in the heat characteristics as seen in Fig. \ref{gammaresults}(b)-(c). We can see a similar effect if we add a finite intrinsic uniaxial anisotropy for the localized spin. As shown in Fig. \ref{anisotropyresults}(a) the intrinsic anisotropy gives different dynamics of the spin due to the interplay with the dynamic exchange fields. This has a major impact on the spin-dependent heat current in Fig. \ref{anisotropyresults}(b) and the net heat flow in Fig. \ref{anisotropyresults}(c). Thus, the local anisotropies of the molecular spin affect the heat currents through the spin dynamics the SMM.

The exchange interaction between the local magnetic moment and the molecular electronic structure, can be tuned by varying the exchange coupling $v$. The magnitude of the exchange coupling $v$, also governs the rate of change of the spin, which is shown in Fig. \ref{exchangeresults}(a). For increasing exchange coupling, the spin-dependent heat current decreases, Fig. \ref{exchangeresults}(d), while net heat transfer increases, Fig. \ref{exchangeresults}(c).
This shows a clear dependence of the coupling on the local magnetic moment. As seen in Fig.  \ref{exchangeresults}(c), there will only be a net heat transfer in the junction when there is a finite exchange coupling, since it is governed by the interaction with the local spin moment. In the same way as for decreasing tunneling coupling, there is an effect on the dynamical local anisotropies of the molecule due to an increased exchange coupling, see Fig. \ref{exchangeresults}(d) (cf. Fig. \ref{gammaresults}(d)).

The results clearly show that there is significant heat transfer in the transient regime and that this can be connected to the localized moment of the SMM and its dynamics. This is in agreement with previous works on time-dependent control of thermoelectric properties \cite{Crepieux2011,Tagani2012,Liu2012a,Zhou2015,Dare2016}. The results presented here show that there are further possibilities of control in the case of dynamic SMMs when including the spin degrees of freedom. Using bias, tunneling and exchange coupling we can tune the characteristics of the dynamic heat transfer. As we are working in the dynamic regime with non-linear equations, we have not considered to calculate the thermoelectric coefficients, although it can easily be done in linear response in the stationary limit as shown in numerous works \cite{Dubi2009a, Swirkowicz2009, Wang2010,  Misiorny2014, Misiorny2015}.

\section{Conclusions}

We have investigated spin-dependent heat signatures in a SMM and its connection to the SMM spin dynamics. We have shown that signatures in the heat current can be attributed to both charge and spin degrees of freedom. The latter can be related to the spin-dependent Peltier effect. Distinct features in the heat flow can be connected to the spin-dependent drive and fluctuations which opens possibilities to engineer thermoelectric devices using driven SMMs. Increasing the bias voltage can introduce a reversal of the net heat transfer which can be attributed to the Zeeman split created by the external magnetic field and localized spin. By tuning the tunneling coupling or the exchange coupling, the dynamic exchange fields and local anisotropies of the SMM can be modulated, and lead to a significant change of the spin-dependent net heat transfer in the system.

\section{Acknowledgments}
The authors thank A. Sisman, P. Oppeneer and R. L\'{o}pez for fruitful discussions. The work is supported by Vetenskapsr\aa det, SNIC 2018/8-29 and Colciencias (The Colombian Department for Science, Technology and Innovation).

\appendix

\section{Lesser/greater quantum Green's function}
\label{app-NEGF}

The lesser/greater forms of the bare GF is given by the Keldysh equation
\begin{align}
\mathbf{g}^{</>}(t,t') = &
\int\mathbf{g}^{r}(t,\tau)\boldsymbol{\Sigma}^{</>}(\tau,\tau')\mathbf{g}^{a}(\tau',t')d\tau d\tau'
\nonumber\\=&
g_{0}^{</>}(t,t')\sigma_{0}+\bfsigma\cdot\bfg_{1}^{</>}(t,t')
,
\end{align}
where
\begin{align}
g_{0}^{</>}(t,t')=&
\int\left(g_{0}^{r}\Sigma_{0}^{</>}g_{0}^{a}+\bfg_{1}^{r}\Sigma_{0}^{</>}\cdot\bfg_{1}^{a}\right.
\nonumber\\&
\left. +g_{0}^{r}\bfSigma_{1}^{</>}\cdot\bfg_{1}^{a}+\bfg_{1}^{r}\cdot\bfSigma_{1}^{</>}g_{0}^{a}\right)d\tau d\tau',
\\
\mathbf{g}_{1}^{</>}(t,t')=&
\int\left(g_{0}^{r}\bfSigma_{1}^{</>}g_{0}^{a}+\bfg_{1}^{r}\cdot\bfSigma_{1}^{</>}\bfg_{1}^{a}\right.
\nonumber\\&
\left. +g_{0}^{r}\Sigma_{0}^{</>}\bfg_{1}^{a}+\bfg_{1}^{r}\Sigma_{0}^{</>}g_{0}^{a}\right)d\tau d\tau'.
\end{align}
Here, we  suppressed the time-dependence of the propagators in the integrands for clarity and 
the retarded/advanced GF are defined as
\begin{subequations}
	\begin{align}
	g_{0}^{r/a}(t,t')=&
	(\mp i)\theta(\pm t\mp t')\sum_{\sigma}e^{-i(\varepsilon_{\sigma}\mp i\Gamma_{\sigma}/2)(t-t')}/2
	,
	\\
	\bfg_{1}^{r/a}(t,t')=&
	(\mp i)\theta(\pm t\mp t')\sum_{\sigma}\sigma_{\sigma\sigma}^{z}e^{-i(\varepsilon_{\sigma}\mp i\Gamma_{\sigma}/2)(t-t')}\mathbf{\hat{z}}/2
	.
	\end{align}
\end{subequations}
Extending this to the lesser/greater forms of the full GF of the system we have
\begin{align}
G_{0}^{</>}(t,t')=&
g_{0}^{</>}(t,t')
-v\int\left(g_{0}^{r}\left\langle \mathbf{S}\right\rangle \mathbf{\cdot g_{1}^{</>}}
+g_{0}^{</>}\left\langle \mathbf{S}\right\rangle \mathbf{\cdot}\mathbf{g}_{1}^{a}
\right.
\nonumber\\&
\left.
+\mathbf{g}_{1}^{r}\cdot\left\langle \mathbf{S}\right\rangle g_{0}^{</>} +\mathbf{g}_{1}^{</>}\cdot\left\langle \mathbf{S}\right\rangle g_{0}^{a}
\right.
\nonumber\\&
\left.
+i\left[\mathbf{g}_{1}^{r}\times\left\langle \mathbf{S}\right\rangle \right]\mathbf{\cdot}\mathbf{g}_{1}^{</>}
+i\left[\mathbf{g}_{1}^{</>}\times\left\langle \mathbf{S}\right\rangle \right]\mathbf{\cdot}\mathbf{g}_{1}^{a}\right)d\tau,
\\
\mathbf{G}_{1}^{</>}(t,t')=&
\mathbf{g}_{1}^{</>}(t,t')
-v\int\left(g_{0}^{r}\left\langle \mathbf{S}\right\rangle g_{0}^{</>}
+g_{0}^{</>}\left\langle \mathbf{S}\right\rangle g_{0}^{a}\right.
\nonumber \\&
+(\textbf{g}_{1}^{r}\cdot\left\langle \mathbf{S}\right\rangle )\textbf{g}_{1}^{</>}
+(\textbf{g}_{1}^{</>}\cdot\left\langle \mathbf{S}\right\rangle ) \textbf{g}_{1}^{a}
\nonumber \\&
+i\left[\mathbf{g}_{1}^{r}\times\left\langle \mathbf{S}\right\rangle \right]g_{0}^{</>}
+i\left[\mathbf{g}_{1}^{</>}\times\left\langle \mathbf{S}\right\rangle \right]g_{0}^{a}
\nonumber \\&
+ig_{0}^{r}\left[\left\langle \mathbf{S}\right\rangle \times\mathbf{g}_{1}^{</>}\right]
+ig_{0}^{</>}\left[\left\langle \mathbf{S}\right\rangle \times\mathbf{g}_{1}^{a}\right]
\nonumber \\&
+\left.i\left[\mathbf{g}_{1}^{r}\times\left\langle \mathbf{S}\right\rangle \right]\mathbf{\times}\mathbf{g}_{1}^{</>}
+i\left[\mathbf{g}_{1}^{</>}\times\left\langle \mathbf{S}\right\rangle \right]\mathbf{\times}\mathbf{g}_{1}^{a}\right)d\tau.
\end{align}

\section{Interactions in terms of Green's functions}
\label{app-Interactions}
The corresponding Heisenberg ($J_{H}$), anisotropic Ising ($\mathbb{I}$) and anisotropic Dzyaloshinsky-Moriya (\textbf{D}) interactions expressed in terms of the charge and spin components of the GFs are
\begin{subequations}
	\label{eq-JID}
	\begin{eqnarray}
	J_{H}(t,t') & = & iev^{2}\theta(t-t')\left(G_{0}^{<}(t',t)G_{0}^{>}(t,t')\right.\nonumber \\
	&  & \left.-G_{0}^{>}(t',t)G_{0}^{<}(t,t')-\mathbf{G}_{1}^{<}(t',t)\cdot\mathbf{G}_{1}^{>}(t,t')\right.\nonumber \\
	&  & \left.+\mathbf{G}_{1}^{>}(t',t)\cdot\mathbf{G}_{1}^{<}(t,t')\right),
	\label{Heisenberg}
	\end{eqnarray}
	\begin{eqnarray}
	\mathbb{I}(t,t') & = & iev^{2}\theta(t-t')\left(\mathbf{G}_{1}^{<}(t',t)\mathbf{G}_{1}^{>}(t,t')\right.\nonumber \\
	&  & \left.-\mathbf{G}_{1}^{>}(t',t)\mathbf{G}_{1}^{<}(t,t')+\left[\mathbf{G}_{1}^{<}(t',t)\mathbf{G}_{1}^{>}(t,t')\right.\right.\nonumber \\
	&  & \left.\left.-\mathbf{G}_{1}^{>}(t',t)\mathbf{G}_{1}^{<}(t,t')\right]^{t}\right),
	\label{Ising}
	\end{eqnarray}
	\begin{eqnarray}
	\mathbf{D}(t,t') & = & -ev^{2}\theta(t-t')\left(G_{0}^{<}(t',t)\mathbf{G}_{1}^{>}(t,t')\right.\nonumber \\
	&  & \left.-G_{0}^{>}(t',t)\mathbf{G}_{1}^{<}(t,t')-\mathbf{G}_{1}^{<}(t',t)G_{0}^{>}(t,t')\right.\nonumber \\
	&  & \left.+\mathbf{G}_{1}^{>}(t',t)G_{0}^{<}(t,t')\right).
	\label{DM}
	\end{eqnarray}
\end{subequations}


\end{document}